\documentclass[sigconf,9pt]{acmart}
\AtBeginDocument{%
  \providecommand\BibTeX{{%
    \normalfont B\kern-0.5em{\scshape i\kern-0.25em b}\kern-0.8em\TeX}}}

\setcopyright{acmlicensed}
\copyrightyear{2024}
\acmYear{2024}
\acmDOI{https://doi.org/10.1145/3649329.3655671}

\acmBooktitle{DAC '24: Design Automation Conference, June 23--27, 2024, San Francisco, CA} 

\acmConference[DAC '24]{Design Automation Conference}{June 23--27, 2024}{San Francisco, CA}

\acmISBN{979-8-4007-0601-1/24/06}

\usepackage{hhline}
\usepackage{colortbl}
\usepackage{centernot}
\usepackage{pbox}
\usepackage{amsmath}
\usepackage{amsfonts}
\usepackage{url}
\usepackage{bm}
\usepackage{comment}
\usepackage{graphicx}
\usepackage{subfigure}
\usepackage{epstopdf}
\usepackage{multirow}
\usepackage{color}
\usepackage{tabu}
\usepackage{mdframed}
\usepackage{syntax}
\usepackage{fancyvrb}

\usepackage[ruled,linesnumbered,algo2e]{algorithm2e}
\usepackage[noend]{algpseudocode}
\usepackage{algorithmicx}
\usepackage{algpseudocode}
\usepackage{paralist}
\usepackage{stmaryrd}
\usepackage{tikz}
\usepackage[referable]{threeparttablex}
\usepackage{tabularx}
\usepackage{booktabs}
\usepackage{array}
\usepackage{fancyhdr}
\usepackage{float}
\usepackage{xspace}
\usepackage{pifont}
\usepackage{balance}
\usepackage{xcolor}

\usepackage{tablefootnote}
\usepackage[normalem]{ulem}

\usepackage{tikz}
\usepackage{xcolor}
\newcommand*\circled[1]{\tikz[baseline=(char.base)]{
            \node[shape=circle,fill,inner sep=1pt,scale=0.8] (char) {\textcolor{white}{#1}};}}

\usepackage{graphicx}
\usepackage{adjustbox}
\def\halfcheckmark{\tikz\draw[scale=0.4,fill=black](0,.35) -- (.25,0) -- (1,.7) -- (.25,.15) -- cycle (0.75,0.2) -- (0.77,0.2)  -- (0.6,0.7) -- cycle;}

\def\fullcheckmark{\tikz\draw[scale=0.4,fill=black](0,.35) -- (.25,0) -- (1,.7) -- (.25,.15);}

\usepackage{float}

\newfloat{figtab}{htb}{fgtb}
\makeatletter
  \newcommand\figcaption{\def\@captype{figure}\caption}
  \newcommand\tabcaption{\def\@captype{table}\caption}
\makeatother

\definecolor{princetonorange}{RGB}{255,143,0}

\definecolor{lightgreen}{RGB}{198, 224, 183}

\definecolor{lightred}{RGB}{240, 205, 176}

\usepackage{geometry}
\begin{document}

\title{\vspace{-.4in}Annotating Slack Directly on Your Verilog:\\Fine-Grained RTL Timing Evaluation for Early Optimization}

\author{Wenji Fang}
\affiliation{%
  \institution{HKUST \& HKUST(GZ)}
  \country{}
}
\email{wenjifang1@ust.hk}

\author{Shang Liu}
\affiliation{%
  \institution{HKUST}
  \country{}
}
\email{sliudx@connect.ust.hk}

\author{Hongce Zhang}
\authornote{Corresponding Author}
\affiliation{%
  \institution{HKUST \& HKUST(GZ)}
  \country{}
}
\email{hongcezh@ust.hk}

\author{Zhiyao Xie}
\authornotemark[1]
\affiliation{%
  \institution{HKUST}
  \country{}
}
\email{eezhiyao@ust.hk}

\pagestyle{empty}

\begin{abstract}

In digital IC design, compared with post-synthesis netlists or layouts, the early register-transfer level (RTL) stage offers greater optimization flexibility for both designers and EDA tools. However, timing information is typically unavailable at this early stage. Some recent machine learning (ML) solutions propose to predict the total negative slack (TNS) and worst negative slack (WNS) of an entire design at the RTL stage, but the fine-grained timing information of individual registers remains unavailable. In this work, we address the unique challenges of RTL timing prediction and introduce our solution named RTL-Timer. To the best of our knowledge, this is the first fine-grained general timing estimator applicable to any given design. RTL-Timer explores multiple promising RTL representations and proposes customized loss functions to capture the maximum arrival time at register endpoints. RTL-Timer's fine-grained predictions are further applied to guide optimization in a standard synthesis flow. The average results on unknown test designs demonstrate a correlation above 0.89, contributing around $3\%$ WNS and $10\%$ TNS improvement after optimization. \looseness=-1

\end{abstract}

\maketitle

\vspace{-.1in}
\section{Introduction}\label{sec:intro}

Performance is a primary design objective in digital integrated circuit (IC) design. To achieve desired performance, huge engineering efforts are spent on the analysis and optimization of \emph{timing}, which describes the maximum delays of signal propagation in IC. However, accurate static timing analysis (STA) tools require precise resistance and capacitance values as inputs, which are often unavailable until late post-layout or sign-off stages.

However, the sign-off stage is often too late to maximally optimize timing. Optimizations are generally preferred at the early stage and high-abstraction level, when many design decisions are not finalized yet. But such early optimization needs a good early timing evaluation to guide it, predicting the ultimate timing in advance. It is extremely challenging for traditional analytical STA tools~\cite{huang2015opentimer} to predict timing at early design stages. For a netlist, due to the lack of wire length information, industry-standard STA tools fail to correlate well with ground-truth timing labels from layout~\cite{xie2022preplacement}. As for the even earlier register-transfer level (RTL), existing STA tools do not even provide any guesses.

The RTL is a critical stage where designers define precise design behaviors with hardware description languages (HDLs) like Verilog, VHDL, or Chisel. Compared with post-synthesis netlists or layouts, the early RTL stage enables significantly higher optimization flexibility, maximally allowing designers or EDA tools to make fine-grained design decisions. However, timing evaluation is unavailable at this early stage and thus optimization lacks guidance.

\begin{figure}[!t]
  \centering
  \includegraphics[width=0.95\linewidth]{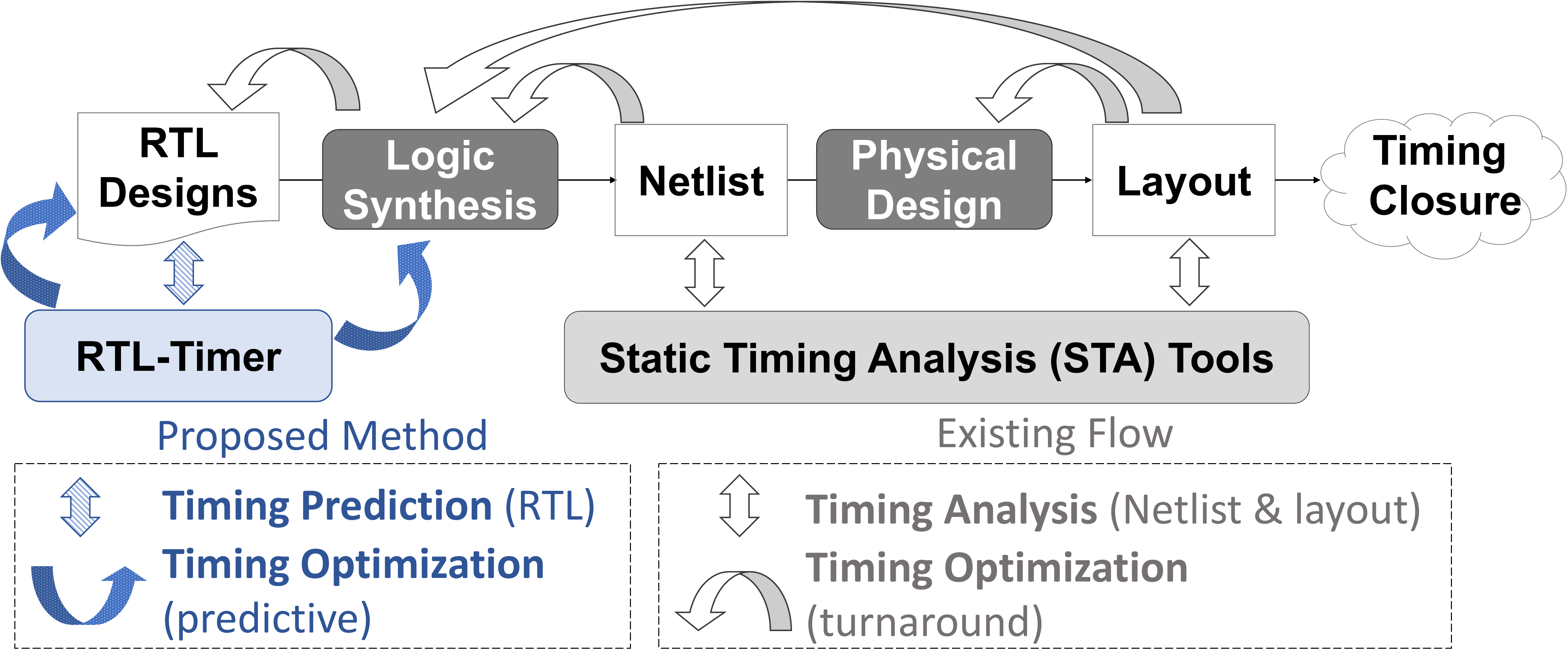}
  \vspace{-.1in}
  \caption{Design flow with our RTL-Timer, enabling RTL-stage timing evaluation for predictive optimizations.\looseness=-1} 
  \label{fig:motivation}
  \vspace{-.2in}
\end{figure}

In recent years, machine learning (ML) methods have been developed to provide early timing predictions. Explorations mostly target the layout~\cite{kahng2018using, barboza2019machine, guo2022timing, cao2022tf, wang2023restructure, he2022accurate} and netlist~\cite{xie2022preplacement} stages. But the more challenging early RTL stage is seldom explored until 2022, largely due to its essentially higher difficulty. We summarize two unique challenges of RTL-stage timing prediction below. They make most post-synthesis solutions~\cite{kahng2018using, barboza2019machine, guo2022timing, cao2022tf, wang2023restructure, he2022accurate, xie2022preplacement} inapplicable: 
\begin{enumerate}
    \item Design RTL is originally in HDL code format, which cannot be directly processed by either ML or traditional STA tools. 
    \item There is no direct mapping between most RTL signals (i.e., model raw input) and post-synthesis cells/nets, where delay labels are supposed to be annotated. 
\end{enumerate}

In this work, we tackle the above two challenges with our solution named RTL-Timer, whose position in the design flow is shown in Fig.~\ref{fig:motivation}. To the best of our knowledge, this is the first general fine-grained RTL-stage timing model that is applicable to any given new design. In addition to overall worst negative slack (WNS) and total negative slack (TNS), our fine-grained solution predicts the slack information of individual sequential RTL signals. This is a highly challenging task, but RTL-Timer is accurate enough to benefit optimization. We have further implemented automatic annotation of predicted fine-grained slack information on user-provided HDL code. We also demonstrated successful optimizations based on RTL-Timer's predicted fine-grained slack information.\looseness=-1

\begin{table}[!t]
\resizebox{0.45\textwidth}{!}{
\begin{tabular}{l||c|c|c|l}
\toprule
\multicolumn{1}{c||}{}                       &   \textbf{Fine-}                  & \multicolumn{2}{c|}{\textbf{General Solution}}        & \multicolumn{1}{c}{}                                                                                                   \\ \cline{3-4}
\multicolumn{1}{c||}{\multirow{-2}{*}{\textbf{Methods}}} & \textbf{Grained} & \textbf{Sequential} & \textbf{Cross-Design} & \multicolumn{1}{c}{\multirow{-2}{*}{\textbf{\begin{tabular}[c]{@{}c@{}}Applied in \\      Optimization\end{tabular}}}} \\ \hline \hline
ICCAD'22~\cite{lopera2022applying}                                               &                                         &                     & \fullcheckmark                    &                                                                                                                        \\
ISCA'22~\cite{xu2022sns}                                                &                                         & \fullcheckmark                  & \fullcheckmark                    &                                                                                                                        \\
ICCAD'22~\cite{sengupta2022good}                                               &                                         & \fullcheckmark                  & \fullcheckmark                    &                                                                                                                        \\
ICCAD'23~\cite{fang2023masterrtl}                                               &                                         & \fullcheckmark                  & \fullcheckmark                    &                                                                                                                        \\
MLCAD'23~\cite{ouyang2023asap}                                               &                                         &                     & \fullcheckmark                    &                                                                                                                        \\
MLCAD'23~\cite{lopera2023using}                                               & \fullcheckmark                                      &                     & \fullcheckmark                    &                                                                                                                        \\
MLCAD'23~\cite{sengupta2023early}                                               & \fullcheckmark                                      & \fullcheckmark              &                   & \multicolumn{1}{c}{\halfcheckmark$^\star$}                                                                        \\
\rowcolor[HTML]{C5D9F1} 
\hline
\textbf{RTL-Timer}      & \fullcheckmark                                      & \fullcheckmark                   & \fullcheckmark                     & \multicolumn{1}{c}{\cellcolor[HTML]{C5D9F1}\fullcheckmark}    \\ \bottomrule                                                        
\end{tabular}
}
\begin{tablenotes}\footnotesize
\item $^\star$ This work~\cite{sengupta2023early} only provides one manual optimization example on a piece of RTL code, without further detailed optimization results on a complete design.
\end{tablenotes} 
\caption{Existing RTL timing evaluators. The earliest exploration started in 2022. RTL-Timer is the first general fine-grained RTL timing estimator, which is also the first to demonstrate a positive impact in realistic optimizations.}
\label{tbl-works}
\vspace{-.3in}
 
\end{table}

\begin{figure*}[!htbp]
    \vspace{-.4in}
  \centering
  \includegraphics[width=0.93\linewidth]{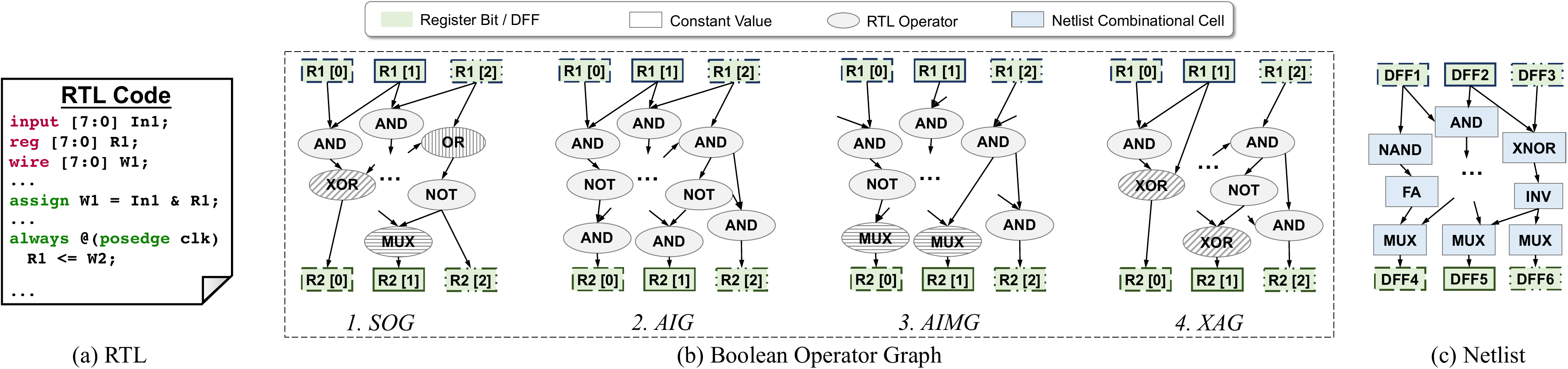}
    \vspace{-.1in}
  \caption{RTL representations explored in this work.}
  \label{fig:repr}
  \vspace{-.1in}
\end{figure*}

RTL-Timer tackles the two aforementioned challenges with innovative techniques. 1) To handle the code format of design RTL, RTL-Timer systematically explores various RTL representations and proposes the ML-friendly ensemble of multiple representation candidates. 2) To address the mismatch between RTL signals (i.e., model raw inputs) and cells/nets with timing labels, we utilize the consistency in registers between RTL and netlist. It captures the maximum arrival time at each register endpoint by sampling multiple paths in its input logic. A customized loss function is developed to enable end-to-end model training. Also, instead of directly evaluating the signals, RTL-Timer focuses on each bit of the signal. Finally, the individual bits are aggregated back into the complete RTL signal, providing a comprehensive evaluation.


Table~\ref{tbl-works} summarizes all existing explorations in RTL-stage timing prediction. Many works~\cite{lopera2022applying, lopera2023using, ouyang2023asap} only accept small-scale combinational designs, while some~\cite{sengupta2023early} cannot even be applied to unknown new designs. They are not general solutions that can apply to any design type. In addition, most methods~\cite{xu2022sns, fang2023masterrtl, sengupta2022good, lopera2022applying, ouyang2023asap} only target TNS or WNS values of a whole design. Also, most works perform prediction only, without being applied in realistic optimizations. Our contributions are summarized below. 

\vspace{-.04in}

\begin{itemize}
\item To the best of our knowledge, RTL-Timer\footnote{It is open-sourced in https://github.com/hkust-zhiyao/RTL-Timer} is the first fine-grained general timing estimator at the early RTL stage. It demonstrates $R=0.89$ and ranking coverage $=80\%$ in fine-grained slack values, and also achieves state-of-the-art accuracy in overall design TNS and WNS predictions.

\item We advance the understanding of data-driven RTL code processing by exploring multiple promising ML-friendly representations and an ensemble of them.

\item To handle the mismatch between RTL signals and netlist cells/nets, we propose a customized ML method to capture the maximum arrival time of each register endpoint. Both regression and learning-to-rank models are explored. 

\item Based on the fine-grained timing prediction, RTL-Timer further predicts overall TNS and WNS, achieving state-of-the-art $R=0.98$ and $R=0.91$, respectively.

\item To demonstrate the effectiveness of RTL-Timer and its benefit in early optimizations, we apply it in two unprecedented applications: 1) We enabled automatic annotating slack prediction of sequential signals in RTL code; 2) We control optimization options \texttt{group\_path} and \texttt{retime} during logic synthesis based on predictions. Experiments show an improvement up to 33.5\% in TNS (avg. 9.9\%) and 16.4\% in WNS (avg. 3.1\%) with negligible impact on power and area. This post-synthesis improvement remains significant after the layout. \looseness=-1


\end{itemize}

\section{Problem Formulation}\label{sec:prob}

We denote an initial design HDL code (e.g., Verilog) as $\mathcal{V}$, and its post-synthesis gate-level netlist as $\mathcal{N}$. Each register as the timing path endpoint\footnote{A tiny portion of endpoints are primary output (PO) pins.} in the design is denoted as $ep_i$. Because of the consistency between RTL sequential signals and netlist registers, $ep_i$ appears in both the RTL design $\mathcal{V}$ and the netlist $\mathcal{N}$. Our framework $F$ will predict the post-synthesis endpoint arrival time\footnote{We assume a fixed clock frequency, implying slack is solely determined by arrival time.\looseness=-1} ($AT_\mathcal{N}$) and the associated ranking ($Rank_\mathcal{N}$), as formulated below:

\textbf{Problem 1} (Register arrival time value prediction). 
\begin{equation}
    {\forall} ep_i \in \mathcal{V}, F_{AT}(ep_i) \rightarrow \ AT_\mathcal{N}(ep_i)
\end{equation}

Note that accurate arrival time prediction at the RTL stage is extremely challenging --- even robust regression models may lead to significant ranking variances. Therefore, we strategically reframe the above as a ranking problem:

\textbf{Problem 2} (Register arrival time ranking prediction). 
\begin{equation}
    {\forall} ep_i \in \mathcal{V }, F_{Rank}(ep_i) \rightarrow \ Rank_\mathcal{N}(ep_i)
\end{equation}

\section{Methodology}\label{sec:method}

\begin{figure*}[!htb]
  \vspace{-.35in}
  \centering
  \includegraphics[width=0.98\linewidth]{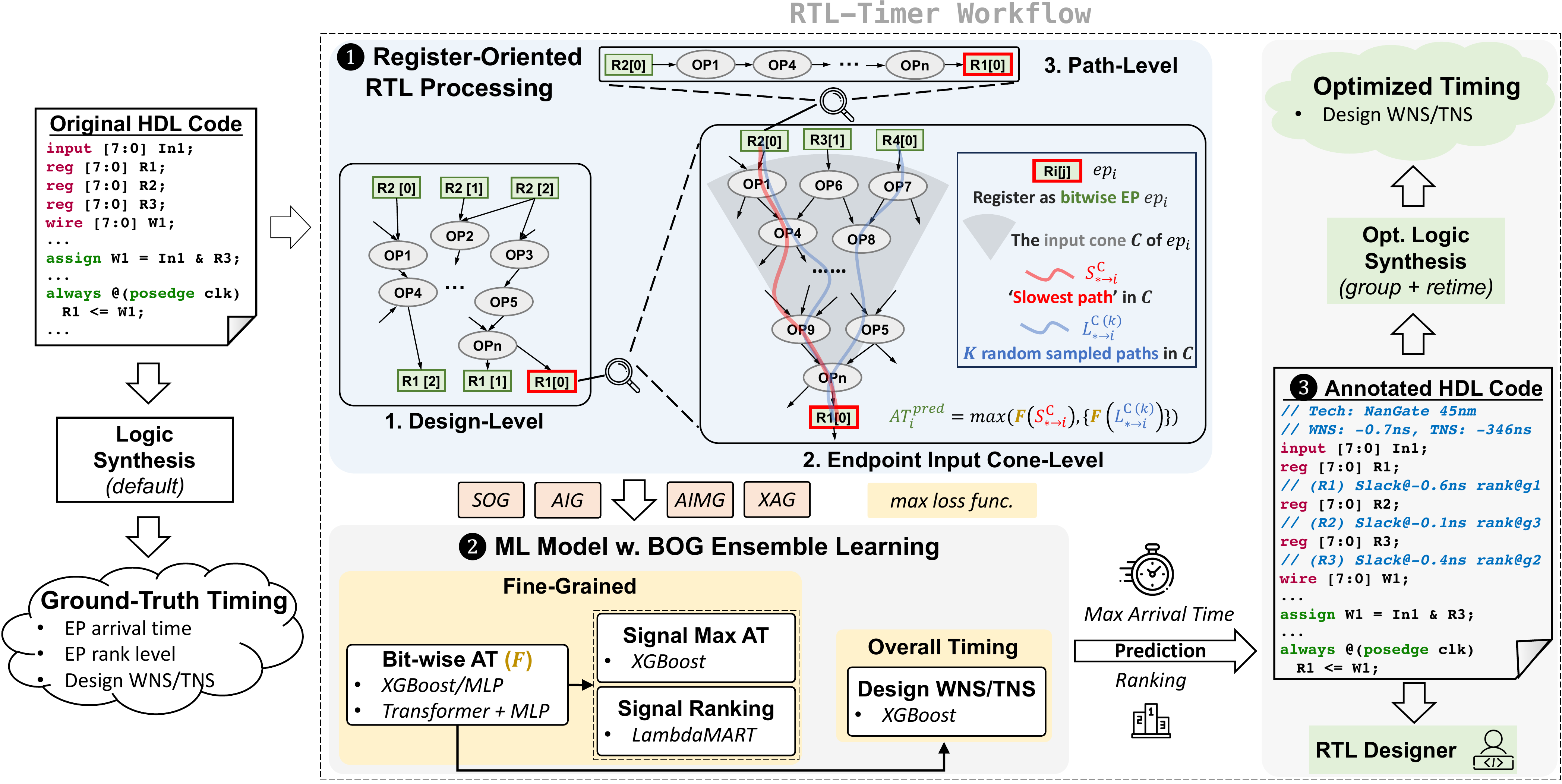}
  \caption{RTL-Timer workflow. We use register-oriented RTL processing and ensemble four RTL representations, enabling fine-grained and overall timing modeling. Predictions are annotated on HDL files, aiding designers and enhancing logic synthesis.\looseness=-1}
  \label{fig:flow}
  \vspace{-.1in}
\end{figure*}


\subsection{Universal ML-friendly RTL Representation}
The first challenge in RTL prediction is that the initial HDL code format cannot be directly processed by the STA tool or ML models. Existing RTL representations, such as Binary Decision Diagrams (BDD), Conjunctive Normal Form (CNF), and And-Inverter Graphs (AIGER), focus primarily on logic transformations for logic synthesis and verification. They are not optimized for ML-based solutions, which require exposing the correlation between RTL and netlist. Existing works have adopted signal-level representations like abstract syntax tree (AST)~\cite{sengupta2022good, xu2022sns} and bit-level representations like simple-operator graph (SOG)~\cite{fang2023masterrtl}. These representations are adopted as ad hoc solutions, without systematically exploring better candidates. \looseness=-1


In this work, we propose a versatile ML-friendly RTL representation, named Boolean operator graph (BOG), denoted as $\mathcal{R}$ and illustrated in Fig.~\ref{fig:repr}(b). BOG is a universal bit-level RTL representation and can be specialized into concrete variants (SOG, AIG, etc.) by selecting different Boolean operators (AND, NOT, OR, XOR, MUX). Besides viewing BOG as a graph of registers and operators, we also treat $\mathcal{R}$ as a pseudo netlist, where registers and operators are viewed as pseudo standard cells from the liberty file.





This bit-level BOG enforces the one-to-one mapping between sequential RTL signal bits and bit-wise netlist registers. It provides the basis for the RTL-stage fine-grained endpoint modeling. 
RTL-Timer employs four distinct BOG representations: SOG, AIG, AIMG, and XAG. 
These four representations share the same functionality for each design, offering rich and multidimensional analytical perspectives. Intuitively, AIG, with only basic AND and NOT operators, provides fundamental insights. In contrast, SOG, consisting of a broader range of operator types, is more similar to the target netlist.
AIMG and XAG reside at the intermediate levels between SOG and AIG. \looseness=-1

Furthermore, we employ ensemble learning to aggregate the strengths of these four representations. This fusion contributes to a more accurate and robust timing modeling approach with significantly reduced variance across various designs. Detailed experimental results will be presented in Section~\ref{sec:expr}. \looseness=-1

\vspace{-.1in}
\subsection{A General Register-Oriented RTL Processing Workflow}
While leveraging the BOG representation $\mathcal{R}$, we face another challenge in RTL prediction: the signals in $\mathcal{R}$ cannot match the cells/nets in the netlist $\mathcal{V}$, leading to a lack of fine-grained labels for $\mathcal{R}$.
Fortunately, the register consistency in BOG enables us to label each bit-wise register (i.e., endpoint $ep_{i}$) with the slack from $\mathcal{N}$ as reported by STA. Nonetheless, a mismatch remains for internal operators.\looseness=-1

Inspired by the propagation mechanism in STA -- where each $ep_{i}$ accumulates arrival time from all its driving registers, and uses the maximum arrival time to compute the slack -- we propose a comprehensive register endpoint-oriented RTL processing approach. This method aims to capture the timing-related pattern of internal operators, as demonstrated in step \circled{1} of Fig.~\ref{fig:flow}. 

If we assume the netlist $\mathcal{N}$ and representation $\mathcal{R}$ were exactly the same, the arrival time of each endpoint $ep_{i}$ in $\mathcal{N}$  will simply only depend on the slowest path $S^{\mathcal{R}}_{* \rightarrow i}$ from $\mathcal{R}$. But logic optimization and technology mapping will optimize $\mathcal{R}$ towards $\mathcal{N}$ in logic synthesis. Therefore, other paths ending at $ep_i$ in $\mathcal{R}$ may also contribute to the ultimate slowest path in $\mathcal{N}$, and we also need to consider these paths, though they may not be the slowest in $\mathcal{R}$. \looseness=-1

Specifically, we backtrack from each $ep_{i}$ to all driving registers in $\mathcal{R}$ to obtain its input cone $C$, which includes all input logic of $ep_{i}$. We then sample two different types of paths from $C$: 1) The slowest path in $\mathcal{R}$: Since we construct $\mathcal{R}$ as a pseudo netlist, we can efficiently traverse $\mathcal{R}$ in topological order and perform the traditional STA algorithm on it. Applying this efficient pseudo-STA process directly on $\mathcal{R}$, we can trace the ``slowest path’’ $S^{\mathcal{R}}_{* \rightarrow i}$ ending at each register $ep_{i}$. 2) Random paths in $\mathcal{R}$: We also randomly sample other paths in $C$, denoted as \{$L^{\mathcal{R}(k)}_{* \rightarrow i} | k \in [1,K_i] $\}, where the sample number $K_i$ is proportional to the number of driving registers. \looseness=-1 

Given a path-level model $F_{AT}$, the ultimate bit-wise max arrival time prediction of $ep_{i}$ depends on the maximum prediction of all paths ending at $ep_{i}$, as formulated below:


\vspace{-.1in}
\begin{equation}
\begin{split}
    AT^{predict}_i&= max(F_{AT}(S^{\mathcal{R}}_{* \rightarrow i}), \{F_{AT}(L^{\mathcal{R}(k)}_{* \rightarrow i})\}),\\
    Loss_i&= LossFunc(AT^{predict}_i, AT^{label}_i).
\end{split}
\end{equation}
\vspace{-.1in}

This loss function is differentiable with respect to model $F_{AT}$, enabling end-to-end gradient-based model training. In this way, we solve the lack of labels resulting from the gap between RTL and netlist and establish a crucial link between the internal operators in $\mathcal{R}$ and the target endpoints' max arrival time.

Building upon bit-wise endpoints, we can calculate the timing for signal-wise endpoints, where the signals are the variables originally defined in the HDL code $\mathcal{V}$. Since an RTL signal can comprise multiple bits, we determine the arrival time of each signal-wise endpoint based on the longest arrival time among all its bits, referred to as the signal max arrival time in subsequent discussions. As demonstrated in Section~\ref{sec:prob}, we address both the signal max arrival time value regression and the critical level ranking tasks. \looseness=-1 

In addition to fine-grained endpoints, we also target the overall timing metrics for the whole design (i.e., TNS and WNS), which are directly calculated utilizing the negative register slack.

\vspace{-.1in}
\subsection{Feature Exploration for RTL Processing}
\begin{figure}[h]
  \begin{minipage}[b]{0.47\linewidth}
  \begin{adjustbox}{width=1\textwidth}
    \begin{tabular}{c||c|c}
    \toprule
    \textbf{Type}           & \textbf{Feature}        &\textbf{Avg. $R$}
    \\ \hline \hline
    \multirow{5}{*}{\textbf{Design}}  & Rank level & \multirow{5}{*}{/}\\
                             &  \% of the endpoint rank&     \\ 
                             &  \# of sequential cells & \\
                             &  \# of combinational cells & \\
                             &  \# of total cells & \\
    
    \hline
    \textbf{Cone}           & \# driving reg of input cone        &0.45\\
    \hline
    \multirow{6}{*}{\textbf{Path}}  & Arrival time by STA on $\mathcal{R}$ &0.43 \\
                             &  \# of level of the timing path & 0.51\\ 
                             &  \# of operators  & 0.56 \\
                             &  Fanout     & 0.40 \\
                             & Load capacitance  & 0.38 \\
                             & Slew  &0.38
               \\ \bottomrule
    \end{tabular}
    \end{adjustbox}
    \tabcaption{Feature summary. }
    \label{tbl:feat}
  \end{minipage} \quad
  \begin{minipage}[b]{0.47\linewidth}
    \centering
    \includegraphics[width=1\linewidth]{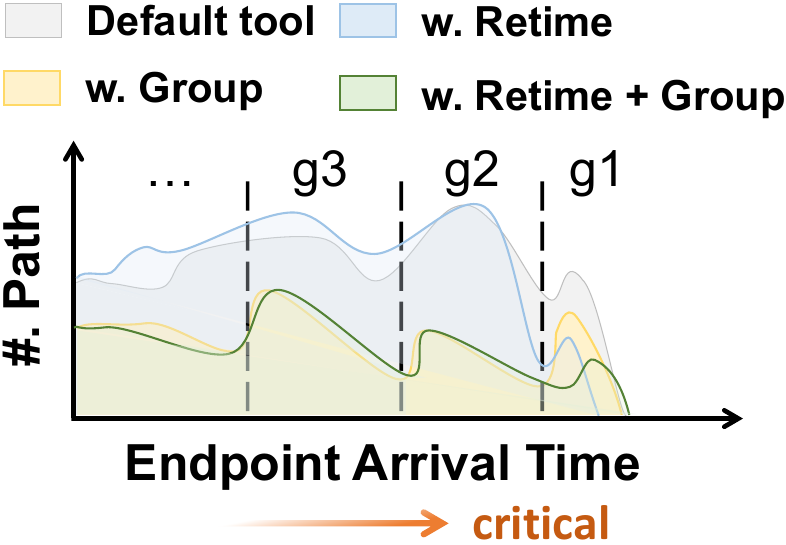}
    \figcaption{Optimization \\options
in logic synthesis.}
    \label{fig:dcopt}
  \end{minipage}%
  
  \vspace{-.2in}
\end{figure}

We extract three levels of features during our RTL processing, as listed in Table~\ref{tbl:feat}.
1) Design-level: Global design features are crucial for comparing endpoints across different designs. Even similar synthesized timing paths from distinct designs can show varied slacks, often due to diverse optimization efforts in logic synthesis. Factors like design size and the critical ranking of timing paths are thus included as our design features. 2) Cone-level: The number of driving registers is used to evaluate the size of the cone. It also helps to differentiate the similar endpoints. 3) Path-level: For timing paths identified by the STA tool, we extract physical-related features and compute key statistics like sum, average, and standard deviation. There is a reasonable correlation $R$ between each feature and the arrival time label at endpoints, providing insightful patterns for further fine-grained model (i.e., $F_{AT}$).
\looseness=-1

Our ensemble approach incorporates the four proposed BOG representations for robust timing modeling. It combines the predictions based on the four BOGs and is supplemented by statistics such as the maximum, minimum, and average of these predictions.

\subsection{ML Model Exploration for RTL Processing}


As for the model, we propose to explore multiple promising ML models to evaluate both the fine-grained timing and TNS/WNS based on our processed RTL representations, as shown in step \circled{2} of Fig.~\ref{fig:flow}.  \looseness=-1

\subsubsection{Bit-wise Endpoint Modeling}

We investigate various ML models, each integrated with the customized loss function tailored for register-oriented RTL processing.  The models include:
1) Tree-based: A lightweight XGBoost model; 2) MLP: A multilayer perceptron (MLP) model; 3) Transformer: This model combines a transformer, for local path modeling, with an MLP to capture global features. 
Once the modeling is complete, we can predict the max arrival time at all bit-wise endpoints within a design, and then further calculate their ranking. \looseness=-1

\subsubsection{Signal-wise Endpoint Modeling}
As previously mentioned, the signal max arrival time is determined by the bit with the longest arrival time. Our signal model is thus constructed leveraging the bit-level predictions.
For the regression model, we employ a lightweight tree-based model.
Regarding the ranking model, we reframe the problem as a learning-to-rank (LTR) task~\cite{cao2007learning}. 
Unlike regression methods that predict absolute target values, LTR learns and predicts the relative ranking among items.
It uses supervised learning with training data comprising queries, each containing a group of documents with features and relevance scores as labels.
In our context, each design is treated as a query, all its endpoints are documents, and their critical ranking levels are the labels. 
The ranking model orders the critical levels of endpoints within each design. We leverage a pair-wise ranking model to capture feature distinctions between timing path pairs. \looseness=-1

\subsubsection{Design Overall Timing Modeling}

Given that TNS and WNS rely on the negative register slack, our proficiency in accurate fine-grained endpoint modeling leads to high prediction accuracy.
In our model, we calculate TNS and WNS using the bit-wise predictions for ensemble features. Additionally, design-level features are incorporated to distinguish among designs of varying scales. Another tree-based model is employed for this regression task.\looseness=-1
\vspace{-.1in}
\subsection{Optimization Enabled by RTL-Timer}
The fine-grained predictions are unprecedentedly applied in two early optimization applications: 1) For manual optimization, RTL-Timer provides early feedback to RTL designers by directly annotating detailed timing information on HDL; 2) For automatic optimization, RTL-Timer can set fine-grained optimization options in the synthesis script. This is illustrated in step \circled{3} of Fig.~\ref{fig:flow}.

\vspace{-.05in}
\subsubsection{Automatic Slack Annotation on HDL}
We have implemented an annotation tool that automatically applies RTL-Timer's timing evaluation onto the original HDL code. It marks the technology node and the overall TNS/WNS for the whole design.  For each sequential signal, it annotates the predicted slack value and its relative ranking group. This tool may become a plug-in for an Integrated Development Environment (IDE), assisting RTL designers to modify timing-critical components without the logic synthesis process.\looseness=-1

\vspace{-.05in}
\subsubsection{Enhancing Logic Synthesis Process.}
RTL-Timer can efficiently control optimization options during logic synthesis. Here, we highlight two key options supported by commercial tools: path grouping and register retiming. By combining these two approaches together, we efficiently improve both TNS and WNS. Fig.\ref{fig:dcopt} illustrates their effects on the arrival time distribution. 

Default logic synthesis tools only focus on the most critical timing violations, often leaving huge space for improvement in other timing endpoints. 
To address this, we divide all endpoints into four groups based on their predicted signal-wise endpoint rankings. We then apply the \texttt{group\_path} command for each register signal, allocating specific optimization efforts to each group. This strategy improves TNS without affecting WNS.
As for retiming, which is not activated by default due to the lack of proper guidance, we focus on the top 5\% of critical endpoints. Utilizing the \texttt{retime} command, registers are repositioned across combinational logic gates for more balanced timing results, particularly beneficial for WNS optimization. \looseness=-1

Note that the specific register names need to be assigned to the above two commands, which was impossible without our fine-grained predictions.\looseness=-1




\section{Experimental Results} \label{sec:expr}
\subsection{Experimental Setup}
We implement our models using Scikit-learn and Pytorch frameworks. Regarding the model hyper-parameters, all the XGBoost models across different granularities are constructed with 100 estimation trees and a maximum depth of 45. The MLPs are configured with 3 layers and a hidden dimension of 512. The transformer model shares the same hyper-parameters as used in~\cite{fang2023masterrtl, xu2022sns}. Since there is no existing work on the general fine-grained timing prediction for sequential RTL designs, we adopt the SOTA layout-stage solution from~\cite{wang2023restructure} as a baseline, where we customize a GNN model to capture the bit-wise endpoint timing information. For the Learning-to-Rank task, we employ the pairwise LambdaMART algorithm with 100 estimators and a maximum depth of 30. 

We train and evaluate our model on 21 open-source RTL designs using 10-fold cross-validation. Training and test datasets include strictly different designs. The benchmark spans various mainstream HDLs, covering a wide design scale range from 6K to 510K gates, as detailed in Table~\ref{tbl:bench}. For dataset generation, we utilize Synopsys Design Compiler and Cadence Innovus with the NanGate 45nm PDK, for logic synthesis and physical design, respectively. Static timing analysis is performed using Synopsys Prime Time. 

\begin{table}[h]
\resizebox{0.43\textwidth}{!}{
\begin{tabular}{l||c|c|c|c}
\toprule
\multicolumn{1}{c||}{\multirow{2}{*}{\textbf{Benchmarks$^\star$}}} & \multirow{2}{*}{\textbf{\#Designs}} & \multicolumn{2}{c|}{\textbf{Design Size Range}}                             & \multicolumn{1}{c}{\multirow{2}{*}{\textbf{HDL Type}}} \\ \cline{3-4}
\multicolumn{1}{c||}{}                                    &                            & \multicolumn{1}{l}{\#K Gates} & \multicolumn{1}{|l|}{\#K Endpoints} & \multicolumn{1}{c}{}                          \\ \hline \hline
ITC'99                                                 & 6                          & 9 - 45                        & 0.4 - 1.3                         & VHDL                                          \\ \hline
OpenCores                                             & 4                          & 6 - 56                        & 0.2 - 3.8                         & Verilog                                       \\ \hline
Chipyard                                                & 3                          & 20 - 32                       & 2.5 - 4.1                         & Chisel                                        \\ \hline
VexRiscv                                                & 8                          & 7 - 510                       & 1.2 - 146                         & SpinalHDL  
\\ \bottomrule
\end{tabular}
}
\begin{tablenotes} \footnotesize
\item $^\star$ Small designs (<5K Gates) and those dominated by huge memory modules are excluded from the original benchmarks.
\end{tablenotes} 
\caption{Benchmark design information.}
\label{tbl:bench}
\vspace{-.4in}
\end{table}

\subsection{Evaluation Metrics}
To evaluate solutions, we employ multiple metrics: the correlation coefficient ($R$), determination coefficient ($R^2$), mean absolute percentage error (MAPE), and critical level ranking coverage (COVR):
\begin{equation*}
\text{MAPE}= \dfrac{1}{n}\sum_{i=1}^{n}\dfrac{|y_i-\hat{y_i}|}{y_i} \times 100\%, \ \
\text{COVR}= \dfrac{1}{m}\sum_{g=1}^{m}\dfrac{\#(S_g \cap \hat{S_g})}{\#S_g} \times 100\%
\end{equation*}
The $\hat{y}$ and $y$ are predictions and labels. In COVR, $S_g$ is the group set categorized by the critical ranking level. In each design, we divide all the signal-wise endpoints into 4 groups: the top 5\% as group1, 5\%-40\% as group2, 40\%-70\% as group3, and the remaining as group4. Notably, the groups are directly used in the subsequent optimizations.
For each group, we calculate the coverage by dividing the number of prediction-label intersections by the label group size.\looseness=-1  

A higher $R$ or $R^2$ and lower MAPE indicate better regression accuracy, and the higher coverage COVR represents
 a more accurate clustering of endpoints into the criticality groups.

\subsection{Modeling Performance}

\begin{table}[]
\resizebox{0.42\textwidth}{!}{
\begin{tabular}{c||l|c|c|c}
\toprule
\textbf{Fine-Grained}   & \textbf{Method}                    & \textbf{$R$} & MAPE (\%) & COVR (\%) \\ \hline \hline
                                            
                                            & Tree-based w/o sample                        &  0.80                              &  26                                      &  59                                      \\
                                            & MLP                                &  0.71                              &   35                                     & 56                                       \\
                                            & MLP w/o sample                               &   0.65                            &    38                                    &  54                                      \\
                                            & Transformer                  &   0.73                             &   35                                     &  57                    \\  & Customized GNN   &   0.25                           & 53                                       &  46       
\\
                      \multirow{-6}{*}{\textbf{Bit-wise}}                      & \cellcolor[HTML]{C5D9F1}RTL-Timer & \cellcolor[HTML]{C5D9F1}0.88   & \cellcolor[HTML]{C5D9F1}12             & \cellcolor[HTML]{C5D9F1}66             \\ \hline

                      & Regression w/o bit-wise       & 0.56      & 28             & 56 \\
& Ranking w/o bit-wise       & /    & /             & 39 \\
                                            & \cellcolor[HTML]{C5D9F1}RTL-Timer (regression)   & \cellcolor[HTML]{C5D9F1}0.89   & \cellcolor[HTML]{C5D9F1}15             & \cellcolor[HTML]{C5D9F1}71             \\
\multirow{-4}{*}{\textbf{Signal-wise}}      & \cellcolor[HTML]{C5D9F1}RTL-Timer (ranking)       & \cellcolor[HTML]{C5D9F1}/      & \cellcolor[HTML]{C5D9F1}/              & \cellcolor[HTML]{C5D9F1}80             \\ 

\bottomrule \bottomrule
\textbf{Overall} & \textbf{Method}                                & \textbf{$R$}                     & \textbf{$R^2$}                     & MAPE (\%)                                       \\ \hline \hline
& SNS~\cite{xu2022sns}                                            & 0.73                           & 0.58                                     & 33   \\
                                            & MasterRTL~\cite{fang2023masterrtl}   & 0.89                           & 0.74                                     & 15                                     \\
\multirow{-3}{*}{\textbf{WNS}}              & \cellcolor[HTML]{C5D9F1}RTL-Timer       & \cellcolor[HTML]{C5D9F1}0.91   & \cellcolor[HTML]{C5D9F1}0.86             &\cellcolor[HTML]{C5D9F1}12                                              \\ \hline   & ICCAD'22~\cite{sengupta2022good}                                       & 0.65                           & 0.32                                     & 42  \\
                                            & MasterRTL~\cite{fang2023masterrtl}                & 0.96                           & 0.94                                     & 34                                       \\
\multirow{-3}{*}{\textbf{TNS}}              
& \cellcolor[HTML]{C5D9F1}RTL-Timer      & \cellcolor[HTML]{C5D9F1}0.98   & \cellcolor[HTML]{C5D9F1}0.97             & \cellcolor[HTML]{C5D9F1}18
\\ \bottomrule
\end{tabular}
}
\caption{Modeling accuracy comparison and ablation study.}
\label{tbl:exprmodel}
\vspace{-.2in}
\end{table}

\begin{table}[]

\resizebox{0.43\textwidth}{!}{
\begin{tabular}{c||l|c|c|c|c|c}
\toprule
\multicolumn{1}{l||}{}                  & \textbf{Metrics} & \textbf{SOG} & \textbf{AIG} & \textbf{AIMG} & \textbf{XAG} & \cellcolor[HTML]{C5D9F1}\textbf{Ensemble} \\ \hline \hline
\multirow{2}{*}{\textbf{Bit-wise}}    & Avg. R            & 0.85         & 0.75         & 0.76          & 0.77         & \cellcolor[HTML]{C5D9F1}\textbf{0.88}     \\
                                      & Std. R            & 0.18         & 0.25         & 0.26          & 0.21         & \cellcolor[HTML]{C5D9F1}\textbf{0.08}     \\ \hline
\multirow{4}{*}{\textbf{Signal-wise}} & Avg. R            & 0.82         & 0.81         & 0.84          & 0.8          & \cellcolor[HTML]{C5D9F1}\textbf{0.89}     \\
                                      & Std. R            & 0.15         & 0.22         & 0.1           & 0.1          & \cellcolor[HTML]{C5D9F1}\textbf{0.06}     \\ \cline{2-7}
                                      & Avg. COVR         & 65           & 71           & 72            & 71           & \cellcolor[HTML]{C5D9F1}\textbf{80}       \\
                                      & Std. COVR         & 18           & 19           & 21            & 21           & \cellcolor[HTML]{C5D9F1}\textbf{8}  
                                      \\ \bottomrule
\end{tabular}
}
 \caption{Comparison of four representation variants and the effect of ensemble learning to reduce variance.}
\label{tbl:expren}
\vspace{-.4in}
\end{table}

\begin{figure*}
  \vspace{-.4in}
  \centering
  \includegraphics[width=0.95\linewidth]{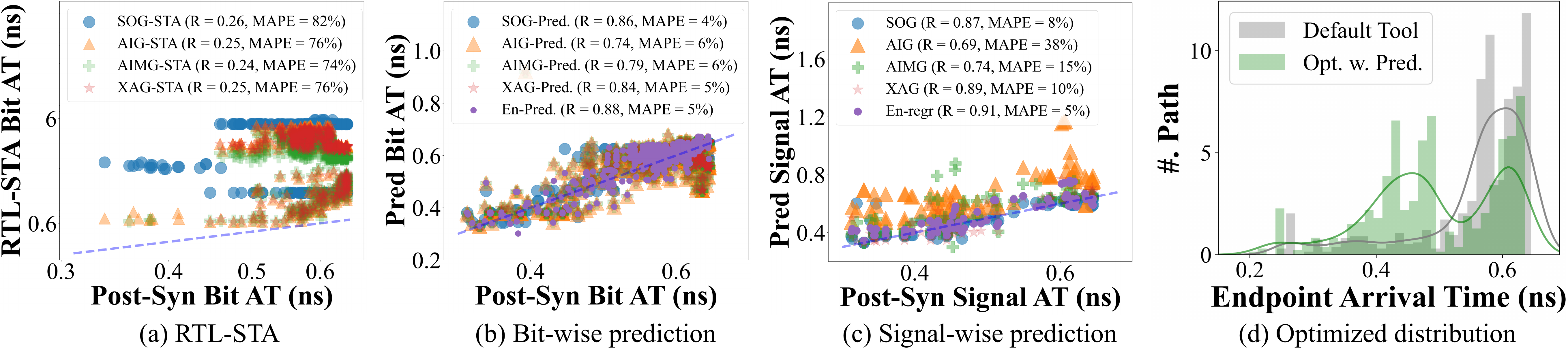}
  \vspace{-.1in}
  \caption{Evaluated/predicted arrival time and optimized distribution for a design example \texttt{b18_1}.}
  \label{fig:scatter}
  \vspace{-.1in}
\end{figure*}

As shown in Table~\ref{tbl:exprmodel}, leveraging the fine-grained bit-wise prediction, which has a correlation of 0.88, our signal-wise prediction achieves a correlation of 0.89 and an 80\% ranking coverage. The lightweight tree-based model outperforms all the deep learning models. We attribute this to our comprehensive feature engineering approach, which treats the original RTL design as tabular data --- a domain where tree-based methods excel~\cite{grinsztajn2022tree}. Additionally, building upon our bit-wise predictions, the overall design TNS and WNS show superior correlations of 0.98 and 0.91, respectively, which outperform all three SOTA methods~\cite{xu2022sns, fang2023masterrtl, sengupta2022good}.

Furthermore, to evaluate the contribution of each strategy of RTL-Timer in fine-grained modelings, we conduct ablation studies by selectively removing key strategies of our solution, also demonstrated in Table~\ref{tbl:exprmodel}:
1) No sampled paths: By exploiting various models with only the slowest path, we observe notable accuracy decreases in all types of models (e.g., $R$ drops 0.08 in the tree-based model). 
2) Removing bit-wise prediction: If we eliminate the detailed analysis at the bit level, and directly model the RTL signal, there will be a significant decrease in accuracy for both regression (i.e., $R$: from 0.89 to 0.56) and ranking (i.e., COVR: from 80\% to 39\%) tasks.
3) Disabling the LTR method: Without the LTR model, the ranking accuracy noticeably diminishes, falling from 80\% to 71\%.\looseness=-1

We also evaluate the four BOG variants and the impact of ensemble learning, as detailed in Table~\ref{tbl:expren}. The results reveal that each representation contributes to the prediction capabilities, with ensemble learning substantially reducing variance, thus ensuring robustness across diverse benchmarks and tasks.

Then we further look into \textit{why ensemble representations are effective}. Analysis of feature importance shows that the average across the four representations forms the core of the final predictions, and omitting any representation leads to a decrease in accuracy. Notably, SOG and AIG carry more weight, reflecting their significant representational differences, whereas AIMG and XAG contribute similarly. Moreover, the inclusion of cone and design features markedly enhances the model's generalization capability across various designs during the representation ensemble.  

Fig.~\ref{fig:scatter} visualizes the experimental results for a design example \texttt{b18_1}. In Fig.~\ref{fig:scatter}(a), the pseudo-STA results of the four representations are shown. While these results do not closely match the post-synthesis arrival time, they offer valuable patterns for further modeling. Utilizing our ML models and ensemble learning technique (denoted as `En'), both bit-wise and signal-wise predictions achieve high accuracy, as depicted in Fig.~\ref{fig:scatter}(b) and (c).

\subsection{Optimization Performance}

Table~\ref{tbl:expropt} demonstrates the automatic optimization results for each design. Leveraging the predicted rankings, our method effectively improves timing on both TNS and WNS for most designs, while maintaining or even decreasing the other design quality metrics (i.e., power and area). Cases where TNS or WNS worsen are considered non-optimized. Practically, designers can concurrently run default and optimization synthesis flows at the same time. In this way, solutions with better outcomes can be selected without time-consuming iterations. `Avg1' considers all the results from the optimization flow, while `Avg2' incorporates default synthesis results for those non-optimized cases.
Experiments illustrate an improvement up to 33.5\% in TNS (avg. 9.9\%) and 16.4\% in WNS (avg. 3.1\%). Notably, compared to the optimization with ground-truth ranking, our approach delivers comparable or superior results.\looseness=-1

Although demonstrating improvements initially in the logic synthesis phase, our optimizations' impact remains significant throughout the placement stages. 
On average, we observe a 4.6\% improvement in WNS and 6\% in TNS after placement. These improvements even persist after the post-placement timing optimization, showing an average enhancement of 3.1\% in WNS and 6.8\% in TNS.\looseness=-1 

Fig.~\ref{fig:scatter}(d) showcases the improved arrival time distribution enabled by the RTL-Timer's predictions. Combining the two optimization options, the original high peak distribution is optimized into two lower peaks with better TNS. Meanwhile, the slowest arrival time (i.e., WNS) is effectively enhanced.  

\vspace{-.08in}

\subsection{Runtime Analysis}
RTL-Timer delivers fast fine-grained timing evaluation, without the need for the logic synthesis process. Overall, the modeling method consumes about 4\% of the default synthesis runtime. It comprises two key parts: 1) RTL process. This involves converting HDL files into BOG variants, a process that can be parallelized. We measure its overhead based on the most time-consuming AIG construction (i.e., 3.2\%). Additionally, the register-oriented RTL processing accounts for 0.9\%. 2) Model inference time. It requires less than 0.1 seconds. 

When employing our optimization in logic synthesis, the runtime extends by an average of 45\%, due to the separate optimization efforts for different path groups and retiming.

\setlength\tabcolsep{1pt}

\begin{table}[]
\begin{adjustbox}{width=0.48\textwidth}
\begin{tabular}{l||ccc||cccc||cccc}
\toprule
\multirow{2}{*}{\textbf{Design}}  & \multicolumn{3}{c||}{\textbf{Singal-wise Pred.}} & \multicolumn{4}{c||}{\textbf{Opt. w. Pred. (\%)}} & \multicolumn{4}{c}{\textbf{Opt. w. Real (\%)}} \\
       & R                 & MAPE       & COVR    & WNS          & TNS          & Pwr      & Area      & WNS            & TNS             & Pwr         & Area        \\ \hline \hline
\texttt{syscdes}   & 0.94              & 26\%             & 94\%              & \cellcolor{lightgreen}-1.3         & \cellcolor{lightgreen}-17          & \cellcolor{lightgreen}-0.8             & \cellcolor{lightgreen}1.8             & \cellcolor{lightgreen}-1.8           & \cellcolor{lightgreen}-17.3           & \cellcolor{lightgreen}0                   & \cellcolor{lightgreen}2.6               \\
\texttt{syscaes}   & 0.86              & 23\%             & 77\%              & \cellcolor{lightgreen}-1.3         & \cellcolor{lightgreen}-13.7        & \cellcolor{lightgreen}2.1              & \cellcolor{lightgreen}3.2             & \cellcolor{lightgreen}-0.1           & \cellcolor{lightgreen}-14.4           & \cellcolor{lightgreen}2.1                 & \cellcolor{lightgreen}3.5               \\
\texttt{Vex_1}    & 0.87              & 24\%             & 86\%              & \cellcolor{lightred}6.9       & \cellcolor{lightred}7         & \cellcolor{lightred}-3.1          & \cellcolor{lightred}-2.8         & \cellcolor{lightgreen}-0.3           & \cellcolor{lightgreen}-1.1            & \cellcolor{lightgreen}0.5                 & \cellcolor{lightgreen}-0.8              \\
\texttt{b20$^\star$}          & 0.91              & 7\%              & 86\%              & \cellcolor{lightred}5.6       & \cellcolor{lightred}-4.3      & \cellcolor{lightred}26.2          & \cellcolor{lightred}25           & \cellcolor{lightred}0.2         & \cellcolor{lightred}-6.6         & \cellcolor{lightred}26.2             & \cellcolor{lightred}23.4           \\
\texttt{Vex_2}   & 0.86              & 16\%             & 83\%              & \cellcolor{lightgreen}-0.2         & \cellcolor{lightgreen}-1.6         & \cellcolor{lightgreen}-0.9             & \cellcolor{lightgreen}0               & \cellcolor{lightgreen}-0.7           & \cellcolor{lightgreen}-1.8            & \cellcolor{lightgreen}-0.6                & \cellcolor{lightgreen}0.3               \\
\texttt{Vex_3}   & 0.93& 30\%             & 86\%              & \cellcolor{lightgreen}-2.8         & \cellcolor{lightgreen}-4.8         & \cellcolor{lightgreen}3.9              & \cellcolor{lightgreen}1.2             & \cellcolor{lightgreen}-0.1           & \cellcolor{lightgreen}-2.2            & \cellcolor{lightgreen}1.9                 & \cellcolor{lightgreen}1                 \\
\texttt{b22$^\star$}         & 0.74              & 18\%             & 83\%              & \cellcolor{lightred}0.7      & \cellcolor{lightred}-4.8      & \cellcolor{lightred}23.4          & \cellcolor{lightred}23.3         & \cellcolor{lightred}2.9         & \cellcolor{lightred}0.4          & \cellcolor{lightred}22.7             & \cellcolor{lightred}20.2           \\
\texttt{b17}        & 0.93              & 8\%              & 75\%              & \cellcolor{lightred}1.9       & \cellcolor{lightred}-5.2      & \cellcolor{lightred}2.2           & \cellcolor{lightred}0            & \cellcolor{lightgreen}-0.9           & \cellcolor{lightgreen}-5.6            & \cellcolor{lightgreen}0.2                 & \cellcolor{lightgreen}1                 \\
\texttt{b17_1}     & 0.94              & 5\%              & 79\%              & \cellcolor{lightred}5.8       & \cellcolor{lightred}-3.2      & \cellcolor{lightred}1.7           & \cellcolor{lightred}2.1          & \cellcolor{lightred}0.9         & \cellcolor{lightred}-5.8         & \cellcolor{lightred}-0.6             & \cellcolor{lightred}0.4            \\
\texttt{Rocket1}     & 0.89              & 11\%             & 63\%              & \cellcolor{lightgreen}-7.1         & \cellcolor{lightgreen}-21.4        & \cellcolor{lightgreen}2.8              & \cellcolor{lightgreen}1.7             & \cellcolor{lightgreen}-3.7           & \cellcolor{lightgreen}-25.4           & \cellcolor{lightgreen}-66                 & \cellcolor{lightgreen}2.9               \\
\texttt{Rocket2}     & 0.92& 18\%             & 64\%              & \cellcolor{lightgreen}-7           & \cellcolor{lightgreen}-23.1        & \cellcolor{lightgreen}-69.4            & \cellcolor{lightgreen}-0.6            & \cellcolor{lightgreen}-4.1           & \cellcolor{lightgreen}-23.1           & \cellcolor{lightgreen}1.6                 & \cellcolor{lightgreen}-0.8              \\
\texttt{Rocket3}     & 0.88              & 12\%             & 69\%              & \cellcolor{lightgreen}-7.2         & \cellcolor{lightgreen}-17.4        & \cellcolor{lightgreen}-69.4            & \cellcolor{lightgreen}-0.6            & \cellcolor{lightgreen}-6.2           & \cellcolor{lightgreen}-18.3           & \cellcolor{lightgreen}-69.7               & \cellcolor{lightgreen}0.8               \\
\texttt{conmax}  & 0.91              & 12\%             & 83\%              & \cellcolor{lightgreen}-1.9         & \cellcolor{lightgreen}-3           & \cellcolor{lightgreen}3.4              & \cellcolor{lightgreen}2.7             & \cellcolor{lightgreen}-0.9           & \cellcolor{lightgreen}-0.6            & \cellcolor{lightgreen}3                   & \cellcolor{lightgreen}2.1               \\
\texttt{b18}         & 0.82& 13\%             & 84\%              & \cellcolor{lightgreen}-16.4        & \cellcolor{lightgreen}-33.5        & \cellcolor{lightgreen}3.8              & \cellcolor{lightgreen}3.9             & \cellcolor{lightgreen}-17.9          & \cellcolor{lightgreen}-35.5           & \cellcolor{lightgreen}3.6                 & \cellcolor{lightgreen}3.4               \\
\texttt{b18_1}      & 0.88              & 10\%             & 86\%              & \cellcolor{lightgreen}-3.9         & \cellcolor{lightgreen}-26          & \cellcolor{lightgreen}0.5              & \cellcolor{lightgreen}-0.4            & \cellcolor{lightgreen}-9.8           & \cellcolor{lightgreen}-27             & \cellcolor{lightgreen}1.8                 & \cellcolor{lightgreen}0.2               \\
\texttt{FPU}         & 0.89              & 31\%             & 85\%              & \cellcolor{lightred}6.2       & \cellcolor{lightred}0.2       & \cellcolor{lightred}0.5           & \cellcolor{lightred}1            & \cellcolor{lightred}4.3         & \cellcolor{lightred}-1.7         & \cellcolor{lightred}-86.6            & \cellcolor{lightred}0.6            \\
\texttt{Marax}  & 0.88& 16\%             & 78\%              & \cellcolor{lightgreen}-1.7         & \cellcolor{lightgreen}-3           & \cellcolor{lightgreen}-0.1             & \cellcolor{lightgreen}-0.1            & \cellcolor{lightred}2.4         & \cellcolor{lightred}-2.2         & \cellcolor{lightred}0                & \cellcolor{lightred}-0.5           \\
\texttt{Vex_4}  & 0.79& 16\%             & 67\%              & \cellcolor{lightgreen}-4.8         & \cellcolor{lightgreen}-18.6        & \cellcolor{lightgreen}0                & \cellcolor{lightgreen}-0.7            & \cellcolor{lightgreen}-7.2           & \cellcolor{lightgreen}-21.9           & \cellcolor{lightgreen}0.4                 & \cellcolor{lightgreen}-1                \\
\texttt{Vex5}  & 0.92              & 4\%              & 81\%              & \cellcolor{lightgreen}-1.8         & \cellcolor{lightgreen}-4.6         & \cellcolor{lightgreen}0.2              & \cellcolor{lightgreen}-1.3            & \cellcolor{lightgreen}-3.6           & \cellcolor{lightgreen}-10.7           & \cellcolor{lightgreen}0.3                 & \cellcolor{lightgreen}-0.6              \\
\texttt{Vex6}  & 0.94& 6\%              & 82\%              & \cellcolor{lightgreen}-1.5         & \cellcolor{lightgreen}-12.8        & \cellcolor{lightgreen}0.4              & \cellcolor{lightgreen}-0.5            & \cellcolor{lightgreen}-3             & \cellcolor{lightgreen}-5.3            & \cellcolor{lightgreen}0.3                 & \cellcolor{lightgreen}-1.3              \\
\texttt{Vex7}  & 0.87              & 6\%              & 81\%              & \cellcolor{lightgreen}-5.7         & \cellcolor{lightgreen}-7.2         & \cellcolor{lightgreen}0.5              & \cellcolor{lightgreen}-0.8            & \cellcolor{lightgreen}-3             & \cellcolor{lightgreen}-12.3           & \cellcolor{lightgreen}0.5                 & \cellcolor{lightgreen}-0.2              \\ \hline
\textbf{Avg1} & \multirow{2}{*}{\textbf{0.89}}     & \multirow{2}{*}{\textbf{15}}    & \multirow{2}{*}{\textbf{80}}  
&\textbf{-1.9}  &\textbf{-10.4} & \textbf{-3.4} & \textbf{2.8} & \textbf{-2.5} &\textbf{-11.2} &\textbf{-7.5} &\textbf{2.7} \\
\textbf{Avg2} &     & & &

\textbf{-3.1}         & \textbf{-9.9}         & \textbf{-5.9}             & \textbf{0.5}             & \textbf{-3}             & \textbf{-10.6}           & \textbf{-5.7}                & \textbf{0.6}
\\
\bottomrule
\end{tabular}
\end{adjustbox}
\begin{tablenotes} \footnotesize
\item $^\star$  Special cases with small size (<15K) and low sequential cell ratio, which are hard to further optimize, resulting in huge power and area overhead.
\end{tablenotes} 
\caption{Optimization enabled by predictions and labels.}
\label{tbl:expropt}
\vspace{-.4in}
\end{table}

\vspace{-.05in}
\section{Conclusion And Future Work}\label{sec:concl}

In this paper, we present RTL-Timer, the first fine-grained general timing estimator for RTL designs, incorporating four RTL representations with a customized loss function to accurately evaluate the arrival time on registers. RTL-Timer facilitates optimizations for both designers and EDA tools. Our future work will focus on enhancing prediction accuracy for more detailed optimization options and explore the potential of automating RTL design optimization using the large language model (LLM).

\section{Acknowledgments}
This work is partially supported by Hong Kong Research Grants Council (RGC) ECS Grant 26208723, National Natural Science Foundation of China (92364102, 62304192), ACCESS – AI Chip Center for Emerging Smart Systems, sponsored by InnoHK funding, Hong Kong SAR, and Guangzhou Municipal Science and Technology Project (Grant No.2023A03J0013).

\bibliographystyle{ACM-Reference-Format}
\bibliography{ref}

\end{document}